\documentclass[times]{qjrms3}

\usepackage[dvips,colorlinks,bookmarksopen,bookmarksnumbered,citecolor=red,urlcolor=red]{hyperref}

\usepackage{moreverb}

\newcommand\um{\mu\mathrm{m}}

\newcommand{\be}{\begin{equation}}
\newcommand{\ee}{\end{equation}}

\def\volumeyear{2007}

\begin{document}

\runningheads{C.~D.~Westbrook}{Sedimentation of small ice crystals}

\title{The fall speeds of sub-100$\mu$m ice crystals}

\author{C.~D.~Westbrook}

\address{Department of Meteorology, University of Reading, UK.}

\corraddr{Dr. Chris Westbrook, Deptartment of Meterology, University of Reading, Earley Gate, Reading, Berkshire, RG6 6BB, UK. E-mail: c.d.westbrook@reading.ac.uk}

\begin{abstract}
Estimates for the sedimentation rate of realistic ice crystals at sizes smaller than 100~$\mu$m are presented. These calculations, which exploit new results for the capacitance of ice crystals, are compared with laboratory studies and found to be in good agreement. The results highlight a weakness in contemporary ice particle fall speed parameterisations for very small crystals, which can lead to sedimentation rates being overestimated by a factor of two. The theoretical approach applied here may also be useful for calculating the sedimentation rate and mobility of non-spherical aerosol particles.

\end{abstract}


\received{5 January 2007}
\revised{\quad}
\accepted{\quad}

\maketitle

\section{Introduction} 
The sedimentation of natural ice particles is fundamental to the evolution of the ice phase in clouds, influencing almost all of the relevant microphysical processes including deposition, riming, aggregation, evaporation and melting. Modelling the sedimentation velocity $v$ of ice crystals is challenging because they are almost universally non-spherical in shape, and span a range of flow regimes. Whereas a rather accurate formula exists for the drag on a sphere (Abraham 1970), theoretical progress for non-spherical particles is more difficult. Importantly, B\"{o}hm (1989) and Mitchell (1996) have attempted to modify Abraham's formulation for an ice particle of arbitrary shape: Mitchell argues that the results are likely to be accurate to within $\sim20\%$ over the whole range of flow regimes experienced by ice particles in the atmosphere. His piecewise fits have recently been refined to provide a continuous power law with variable co-efficients (Khvorostyanov and Curry 2002), and further adjustments have been made for turbulence at high Reynolds number (Mitchell and Heymsfield 2005, Khvorostyanov and Curry 2005). This method has proved rather successful, providing fall velocities consistent with observations for a range of size and shape particles, and the framework itself is very appealing because of its generality. However there is a need to test it, particularly for very small crystals where the boundary-layer ideology underpinning it breaks down. 

Ice crystals smaller than $100\um$ can play an important role in clouds. Such crystal sizes are common at the top of stratiform clouds and Rauber and Tokay (1991) suggest that it is this feature which allows supercooled liquid droplets to persist there in a weak updraught, despite the flux of vapour to the ice crystals. The sedimentation of the crystals is key to a quantitative understanding of this situation, since the longer the crystals reside in the mixed-phase layer, the more time they spend growing at the droplets' expense. This range of crystal sizes is also typical of supercooled fogs (eg. Yagi 1970) and diamond dust. Aircraft contrails are composed of crystals of order $10\um$ in size, and their sedimentation influences the extent to which the contrail removes moisture from the upper troposphere, and could affect its persistence/transition to cirrus. In natural cirrus, there remains much controversy over the concentrations of ice crystals smaller than around $60\um$. Aircraft measurements have indicated that these tiny crystals may exist in very large numbers (see Heymsfield and McFarquhar 2002) and could dominate the optical properties of the cloud; on the other hand, there is also evidence to suggest that some of the measurements of this small crystal mode are an artefact of larger crystals shattering on the probe inlet (Field \etal~2003, McFarquhar \etal~2007, Heymsfield 2007). Doppler lidar observations could inform this debate, since if these tiny crystals were to dominate the optical properties of the cloud, they should also dominate the lidar backscatter and Doppler velocity. Given accurate estimates of the sedimentation rates of these small ice crystals, there is an opportunity to compare these values with the measured frequency distribution of lidar Doppler velocities in cirrus clouds, and to determine whether or not sub-60$\mu$m crystals genuinely have a significant impact on the optical properties of cirrus.

The aim of this short paper is to investigate the sedimentation velocity of small ice crystals where the air flow is dominated by viscous forces. New estimates of the `capacitance' for realistic ice crystal shapes (Westbrook \etal~2008) allows the application of two theoretical results from the physics literature (Roscoe 1949, Hubbard and Douglas 1993) to calculate crystal fall speeds; these estimates are then compared to experimental data, and to the fall-speed parameterisations of Mitchell (1996), Mitchell and Heymsfield (2005) and Khvorstyanov and Curry (2002, 2005); the latter four studies are collectively referred to as MHKC from now on. Upper and lower bounds are also constructed where possible using analytical results for spheres and spheroids, providing further insight.

\section{Viscous sedimentation speeds}
When the Reynolds number $\mathrm{Re}=vD/\nu_k$ is small, viscous forces dominate the flow. Here $D$ is taken to be the maximum dimension of the crystal, and $\nu_k$ is the kinematic viscosity of air. The Stokes solution for a sphere in a viscous flow is well known:
\begin{equation}
v=\left(\frac{g}{6\pi\eta}\right)\times\frac{m}{R}
\label{stokes}
\end{equation}
where $m$ and $R$ are the particle mass and radius respectively, $g$ is the acceleration due to gravity, and $\eta$ is the viscosity of the air. Dimensional analysis suggests that an identical formula should exist for non-spherical ice particles falling in a given orientation (or ensemble of orientations), but with an `effective' hydrodynamic radius in place of the sphere radius. The problem then is to determine this hydrodynamic radius for natural ice crystals.

Very small crystals are affected by rotational Brownian motion and often have an approximately random orientation in the absence of an electric field (eg. Foster and Hallett 2002). Hubbard and Douglas (1993) have provided a theoretical treatment for the average viscous drag on a particle which experiences all possible orientations. They preaverage the Oseen tensor (which acts as a Green's function for the Navier-Stokes equations) over all angles, and find that it is identical to the Green's function for Laplace's equation (to within a constant factor). Based on this observation they find that the angle-averaged flow system can be described through a scalar potential which is fixed on the surface of the settling particle, and zero far from it. This situation is directly analogous to diffusion and electrostatic problems. An overview of their analysis is given in appendix A, but the end result of this approximation is that the hydrodynamic radius for the particle is:
\begin{equation} 
R=C
\label{hubbard}
\end{equation}
where $C$ is the capacitance of the particle in length units. The capacitance is commonly used to estimate the growth/evaporation of ice particles when the mass flux is limited by diffusion of vapour onto the ice surface, and it has recently become possible to calculate this parameter accurately using a Monte Carlo method (Westbrook \etal~2008). The link between molecular diffusion and viscous drag seems sensible since the drag in viscous flow is essentially the product of momentum diffusing away from the falling crystal. Inserting the capacitance for a spheroid into equation \ref{hubbard} yields the exact Perrin (1934, 1936) formula; comparison with experimental data on non-spherical shapes has yielded agreement to within a few percent (Hubbard and Douglas 1993). Blawzdziewicz \etal~(2005) has provided further theoretical arguments to explain why this approximation turns out to be rather accurate. Although there is an increasing torque to orient the ice crystal as the Reynolds number becomes larger, we might expect that the randomly oriented approximation may still be acceptable in a number of cases, especially for particles which do not have strong symmetry (eg. polycrystals) or column/needle crystals where the orienting torque is relatively weak (Katz 1998).

The strongest orientational torque is likely to be for plate-like particles which are observed to orient approximately horizontal if the crystal grows large enough. For example Katz (1998) estimated that a disc with a diameter of $60\um$ and a thickness of $3\um$ would orient approximately horizontally with an angular dispersion of $0.6^{\circ}$. In this situation the angle-averaged Hubbard-Douglas approximation is likely to underestimate the drag. However, Roscoe (1949) provides an approximation for the drag on a horizontally oriented particle in viscous flow. For a thin planar crystal, the Navier-Stokes equations can be transformed into Laplace's equation with a fixed potential at the particle surface. Again, this is directly analagous to the vapour diffusion problem and the hydrodynamic radius is again simply related to the capacitance:
\begin{equation} 
R=\frac{4}{3}C.
\label{roscoe}
\end{equation}
The derivation is straightforward, and is reproduced in appendix B. Strictly this result applies only to planar particles of zero thickness; however comparison between the exact formula for a thin oblate spheroid with aspect ratio 0.1 (Happel and Brenner 1965), and equation \ref{roscoe} with the associated capacitance ($C=0.338D$ Pruppacher and Klett 1997, page 547) reveals a difference of less than 0.5\%. 

The MHKC parameterisations all collapse to the same formula in the small Reynolds number limit, and predict a hydrodynamic radius proportional to the ratio of the projected area of the particle $A$ to its maximum dimension $D$:
\begin{equation}
R=\frac{C_0\delta_0^2}{24}\times\frac{2A}{\pi D}
\label{MHKC}
\end{equation}
where $C_0$, $\delta_0$ are dimensionless parameters characterising the boundary layer shape and thickness (which have no physical meaning in a viscous flow). For liquid drops the parameters suggested by Abraham are used ($C_0=0.292$, $\delta_0=9.06$) which matches the Stokes formula for a spherical particle (substituting $A=\pi R^2$, $D=2R$). For ice particles, MHKC prefer $C_0=0.6$, $\delta_0=5.83$, giving a fall velocity $\sim15\%$ higher than the Stokes result for a spherical particle\footnote{Khvorostyanov and Curry used the smooth sphere parameters for ice particles in their 2002 study, but their subsequent paper (Khvorostyanov and Curry 2005) uses $C_0=0.6$, $\delta_0=5.83$, as do Mitchell (1996) and Mitchell and Heymsfield (2005). In this paper we will assume these latter parameters.}. Note that (\ref{MHKC}) may also be expressed as $R=(C_0\delta_0^2/24)\times \gamma D/2$ where $\gamma$ is the `area ratio' of the particle (ie. the projected area divided by the area of an enclosing circle). 

\subsection{Planar crystals}
Planar crystals are a frequent occurence in the atmosphere. The standard crystal habit diagram suggests these crystals form at temperatures between approximately $0$ to $-3^{\circ}$C and $-8^\circ\mathrm{C}$ to $-25^\circ\mathrm{C}$; however recent laboratory experiments by Bailey and Hallett (2002, 2004), along with in-situ evidence cited in their papers, indicates that this diagram is probably biased by the use of silver iodide in laboratory studies, and plate-like crystals can form at temperatures as cold as $-70^{\circ}$C, particularly at low supersaturations.

\subsubsection{Hexagonal plates}
\begin{figure}
\centering
\includegraphics[width=3in]{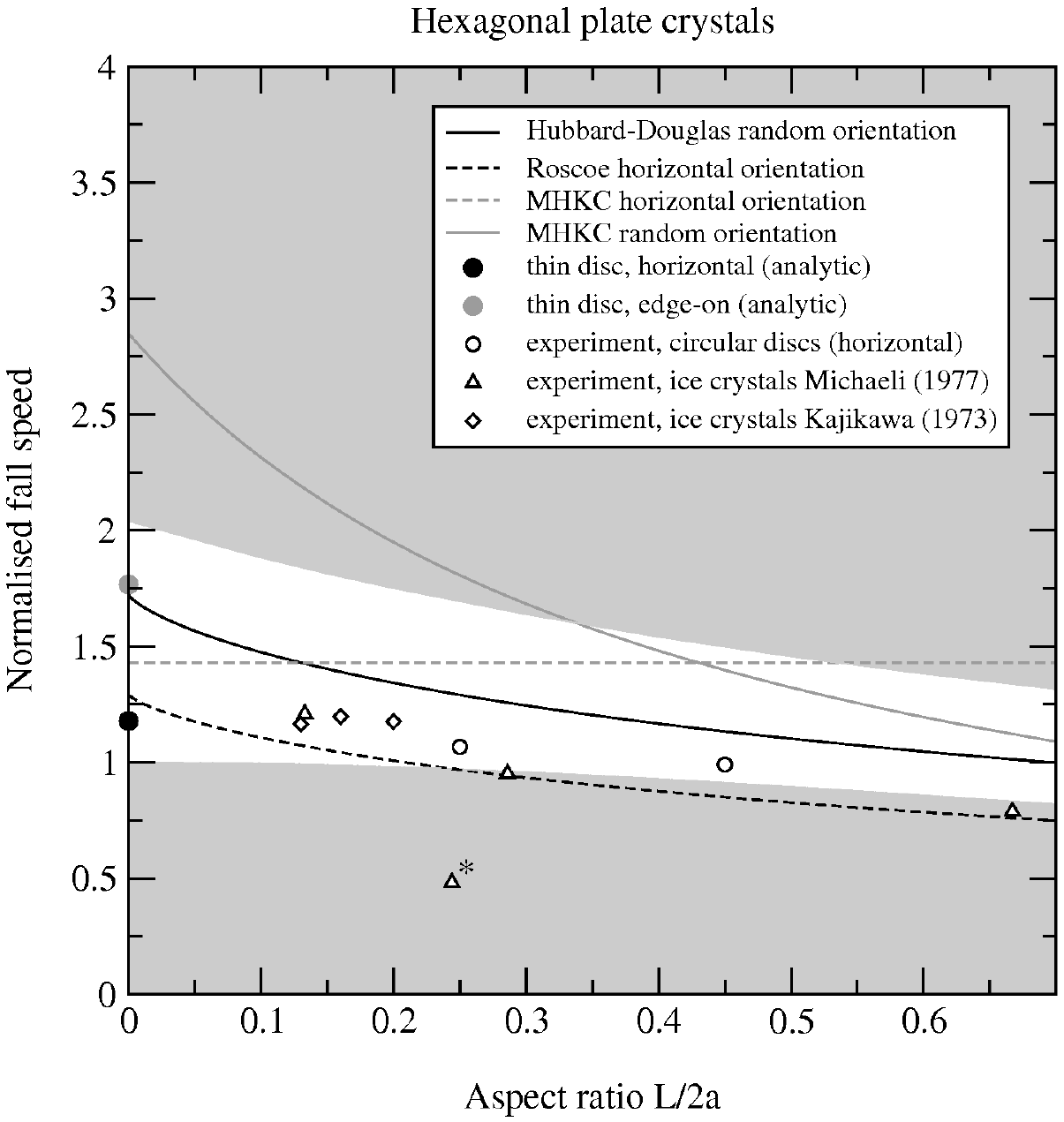}
\caption{\label{plates} Fall speed of hexagonal plate ice crystals, normalised by the fall speed of a sphere with the same mass and maximum dimension. Solid black line shows Hubbard \& Douglas prediction for randomly oriented crystals ($R=C$), dashed black line shows the Roscoe prediction for horizontally oriented crystals ($R=\frac{4}{3}C$). Diamonds and triangles show fall speeds of ice crystals as measured by Kajikawa (1973) and Michaeli (1977) respectively. Asterisk indicates the aspect ratio for this crystal was not measured (see text). Hollow circles are measurements of circular discs falling horizontally. Filled circles at $L/2a=0$ represent analytical predictions for discs of zero thickness falling horizontally (black) and edge-on (grey). MHKC predictions are shown by the grey solid line (randomly oriented) and grey dashed line (horizontally oriented). Shading indicates fall velocities which lie outside the upper and lower bounds derived from analytical results (see text).}
\end{figure}

The capacitance of a hexagonal plate is given by Westbrook \etal~(2008):
 \begin{equation}
 C=0.58a\times[1+0.95(L/2a)^{0.75}]
 \label{cplate}
 \end{equation}
 for a crystal of span $(2a)$ across the basal face, and thickness $L$, with aspect ratio $L/2a<1$. The fall speeds of crystals with aspect ratios in the range 0--0.7 have been calculated using the Hubbard-Douglas and Roscoe approximations: these curves are shown in figure \ref{plates}. To facilitate the comparison between theory and experiment the fall speeds have been normalised relative to the fall speed of a sphere with the same mass and maximum dimension (ie. normalised velocity$=D/2R$). 

Experimental data comes from Michaeli (1977) and Kajikawa (1973). Michaeli measured the mass, dimensions, and sedimentation velocities of free-falling hexagonal plate crystals in a cloud chamber with $D$ ranging from $50$ to $100\um$. From this information $R$ and the normalised velocity were calculated directly: see triangle data points in figure \ref{plates}. The experimental data lies between the Roscoe and Hubbard-Douglas curves, somewhat closer to the former, suggesting the crystals were to a large extent oriented in the horizontal. The point marked by an asterisk appears to be an outlier; note that for this particular point the thickness of the crystal was not directly measured, but calculated using an empirical relationship from a different study - the measured velocity would be more consist with an aspect ratio closer to unity. Similarly, Kajikawa (1973) seeded a supercooled fog in a cold room and measured the fall speeds of the resulting crystals ($15-100\um$)\footnote{Kajikawa's $20\um$ bin was not used here because the fall speeds were measured to be higher than that of an inscribed disc falling edge-on, suggesting that the melted diameter of these tiny crystal is likely have been underestimated.}. Fall streak photographs were used to calculate the velocities, and mass and thickness measurements were taken from similar crystals of the same type and diameter which fell out at the same time. There is a large amount of scatter in his data; however, taking the bin averages from his table 1 and comparing them with his calculated values for circular discs the associated normalised velocties can be estimated (diamonds). The aspect ratios were estimated from his figure 5. Again the data points lie between the Roscoe and Hubbard-Douglas curves indicating horizontal orientation with a certain amount of flutter. 

Jayaweera and Ryan (1972) also measured the fall speed of small plate-like crystals; however, unlike their measurements of columnar crystals (see section 2.2) they did not explicitly record the crystal aspect ratio. Their fit corresponds to a normalised fall speed of 1.41; assuming an aspect ratio of $\simeq0.1$ this value lies between the Roscoe and Hubbard-Douglas curves.

Also shown in figure \ref{plates} are analytical results for infinitely thin circular discs, and experimental data for discs of finite thickness (see Clift \etal~1978). These discs have the same diameter $D$ as the hexagonal plates. For a thin disc falling horizontally the drag is slightly larger than for the hexagonal plate prediction, with the disc falling 9\% slower than the plate. For larger aspect ratios the horizontal discs fall slightly faster than the plates, indicating that the Roscoe curve underestimates $v$ somewhat as we depart from the zero thickness approximation. For a thin circular disc falling edge-on the normalised velocity is slightly higher than the Hubbard-Douglas prediction for a randomly oriented hexagonal plate, which seems sensible. 

The MHKC formula was used to calculate the grey curves in figure \ref{plates} assuming both randomly oriented ($A=a^2[1.3+3(L/2a)]$) and horizontally oriented crystals ($A=2.6a^2$). The former assumption is recommended by Mitchell (1996) for crystals with $\mathrm{Re}<10$. The curve for randomly oriented crystals predicts fall speeds approximately 70--80\% higher than the experimental data for aspect ratios between 0.1--0.3 where most of the data points are clustered. The curve for horizontal orientation gives more realistic fall speeds for thin plates, around 10--20\% higher than the experimental data. The Jayaweera and Ryan data fit is in approximate agreement with the horizontal prediction if the aspect ratio of the crystals is assumed to be around 0.1. The strong sensitivity of eqution \ref{MHKC} to crystal orientation appears to be unphysical given the relatively weak dependence on fall attitude observed for circular discs (for very thin discs edge-on is 45\% faster than for horizontal orientation, a dependence that becomes weaker for increasing aspect ratio, see Clift \etal~1978).

Hill and Power (1956) showed that bounds could be constructed for the drag on a particle of arbitrary shape by using known results for inscribed and circumscribed spheroids. In this vein the fall speed for  an oblate spheroid inscribing the hexagonal plate crystal was calculated using Oberbeck's formula (Happel and Brenner 1965). The edge-on orientation was identified as that corresponding to the maximum possible fall velocity. Any values of $v$ larger than this upper bound must therefore be erroneous, and this indicated by grey shading in figure \ref{plates}. The MHKC curve for randomly oriented crystals violates this bound for aspect ratios less than 0.3; similarly the horizontally oriented curve violates the bound for aspect ratios larger than 0.55. The Hubbard-Douglas and Roscoe curves both lie below the upper bound for all aspect ratios. A second bound may also be constructed (not shown in figure \ref{plates} for clarity) for plates which are \textit{assumed} to lie in a horizontal orientation. The MHKC curve for horizontal plates lies above this upper bound by 10-20\% for all aspect ratios; the Roscoe curve lies below it for all aspect ratios. 

It is also possible to construct a lower bound. Using the drag for a circumscribed sphere (diameter $=\sqrt{L^2+(2a)^2}$) such a bound for was constructed: value lower than this bound are shaded in figure \ref{plates}. The Roscoe curve dips slightly below this lower bound for aspect ratios larger than 0.25, as do a couple of the experimental data points (by a couple of percent). For an aspect ratio of 0.7 the Roscoe curve is 8\% lower than the enclosing sphere lower bound. The suggestion is that the zero-thickness Roscoe approximation progressively underestimates the fall speeds for increasing aspect ratios. The Hubbard-Douglas and MHKC curves lie well above the lower bound.

As the crystals grow larger they may start to deviate slightly from the viscous drag regime, and this could potentially affect some of the experimental data. Keller \etal~(1967) have shown that in this case the viscous drag calculations underestimate the true drag, and therefore the corresponding normalised velocity is reduced. Experimental and numerical results for spheres and discs suggest that for $\mathrm{Re}<1$ this deviation is less than 10\% (Clift \etal~1978). The experimental data for ice crystals all had Reynolds numbers $\mathrm{Re}<0.4$.

Note that there is a slight subtlety in the meaning of the maximum dimension $D$ in the above analysis: for planar crystals we take $D$ to mean the maximum dimension across the basal face ie. $D=2a$ since this is the dimension which is usually measured in observations. If $D$ is taken to mean the three dimensional maximum span ($=\sqrt{L^2+(2a)^2}$) then all the normalised velocities are shifted up by a factor $f=\sqrt{1+(L/2a)^2}$ except for the MHKC curves which are shifted up by a factor $f^2$ since $D$ also appears in the definition of $R$ via equation \ref{MHKC}. The result of this transformation is that the entire MHKC curve for randomly oriented crystals now lies above the upper bound derived above, as do the MHKC predictions for horizontally oriented plates with $L/2a>0.25$; the differences with the experimental data points are magnified by a factor $f$. 

\subsubsection{Branched \& dendritic crystals}
\begin{figure}
\centering
\includegraphics[width=3in]{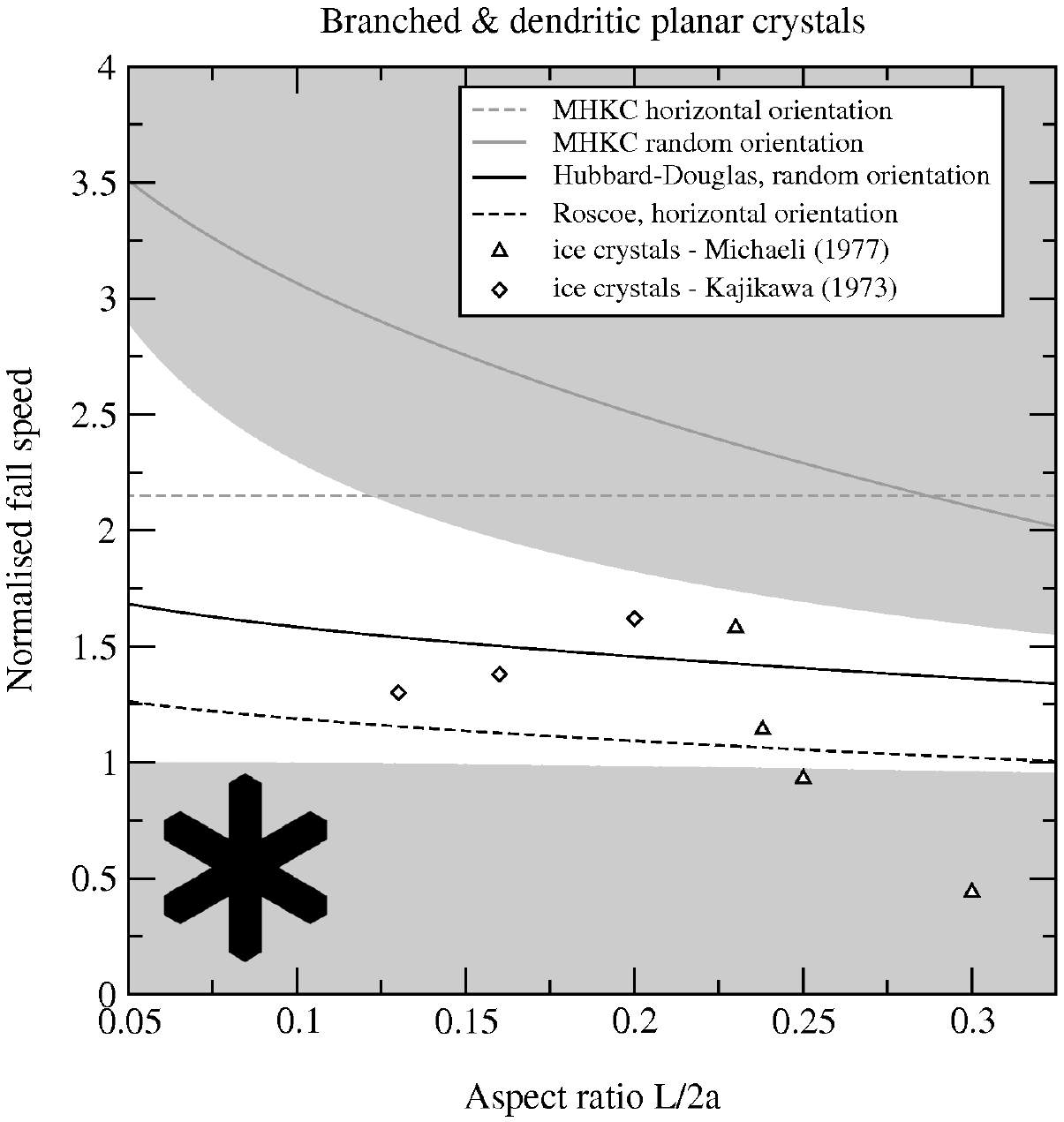}
\caption{\label{dendrites} Normalised fall speed of branched and dendritic planar ice crystals. Solid black line shows Hubbard-Douglas prediction for randomly oriented crystals ($R=C$), black dashed line shows the Roscoe prediction for horizontally oriented crystals ($R=\frac{4}{3}C$). Diamonds and triangles show fall speeds of ice crystals as measured by Kajikawa (1973) and Michaeli (1977) respectively. MHKC predictions are shown by the grey solid (randomly oriented) and grey dashed line (horizontally oriented). Inset is the branched crystal model used in the capacitance and projected area calculations. Shading indicates fall velocities which lie outside the upper and lower bounds derived from analytical results (see text).}
\end{figure}

Kajikawa (1973) and Michaeli (1977) also made measurements of the fall speed of branched and dendritic crystals, and their experimental results are shown in figure \ref{dendrites}. Based on photographs in Kajikawa's paper the capacitance of a `representative' branched crystal has been calculated for aspect ratios between $L/2a=0.05$--$0.35$ using the method outlined in Westbrook \etal~(2008). The outline of the model crystal is shown inset in figure \ref{dendrites}. The width of the branches is 35\% of their length from the crystal centre. The overall span from tip to tip across the crystal is defined as $(2a)$ and the thickness as $L$ in the same way as for the hexagonal plate crystals. The values of $C$ were only marginally lower than for a solid hexagonal plate. The capacitance data were substituted into equations \ref{roscoe} and \ref{hubbard}, and curves fitted to the resulting fall speed predictions: these are shown in figure \ref{dendrites}. The maximum dimension was defined as the span from tip to tip across the projection shown in figure \ref{dendrites}, ie. $D=2a$.

The experimental data appear to be in broad agreement with the range of velocities indicated by the Roscoe and Hubbard-Douglas theories, although the scatter is rather wider than for the hexagonal plate data, probably indicating the range of crystal shapes and perhaps a broader distribution of crystal orientations. The aspect ratio of the crystals observed by Michaeli (triangles) were not directly measured but estimated using an empirical relationship from a different study, further contributing to the experimental scatter.

Also shown in figure \ref{dendrites} are the MHKC predictions for the model crystal described above. Under the horizontal orientation assumption the MHKC curve is around 45\% higher than the experimental data, whilst for the randomly oriented assumption the MHKC curve is around 80\% higher. This is particularly an issue for crystals with thin branches and dendritic features where the projected area $A$ becomes very small but the capacitance is only marginally reduced relative to a solid hexagonal plate (typically by only $\sim$25\% even for quite thin branches, see Westbrook \etal~2008). If the capacitance is the appropriate length scale then this implies that MHKC theory will increasingly overestimate the crystal fall speed as the crystals become more tenuous.

It was not possible to construct an upper bound using an inscribed oblate spheroid as before, because of the branched nature of the crystal. However, Weinberger (1972) showed that a sphere with equal volume acts as an upper bound for the settling velocity of a non-spherical particle in a viscous flow. The normalised velocity for a sphere with the same volume as the model crystal was calculated as shown by the dark grey region in figure \ref{dendrites}. The MHKC curve for random orientation lies entirely above this upper bound. The curve for horizontal orientation also violates the upper bound for aspect ratios larger than 1/8. The Roscoe and Hubbard-Douglas approximations lie below the upper bound for all aspect ratios.

A lower bound may be constructed in the same way as for plate crystals and this is indicated by the shaded region in figure \ref{dendrites}. A couple of the experimental data points from Michaeli fall below this bound, although given the uncertainty in the aspect ratio this could be an artefact for the crystal plotted at $L/2a=0.25$. All of the theoretical curves lie above the lower bound.

\subsection{Columnar ice crystals}
\begin{figure}
\centering
\includegraphics[width=3in]{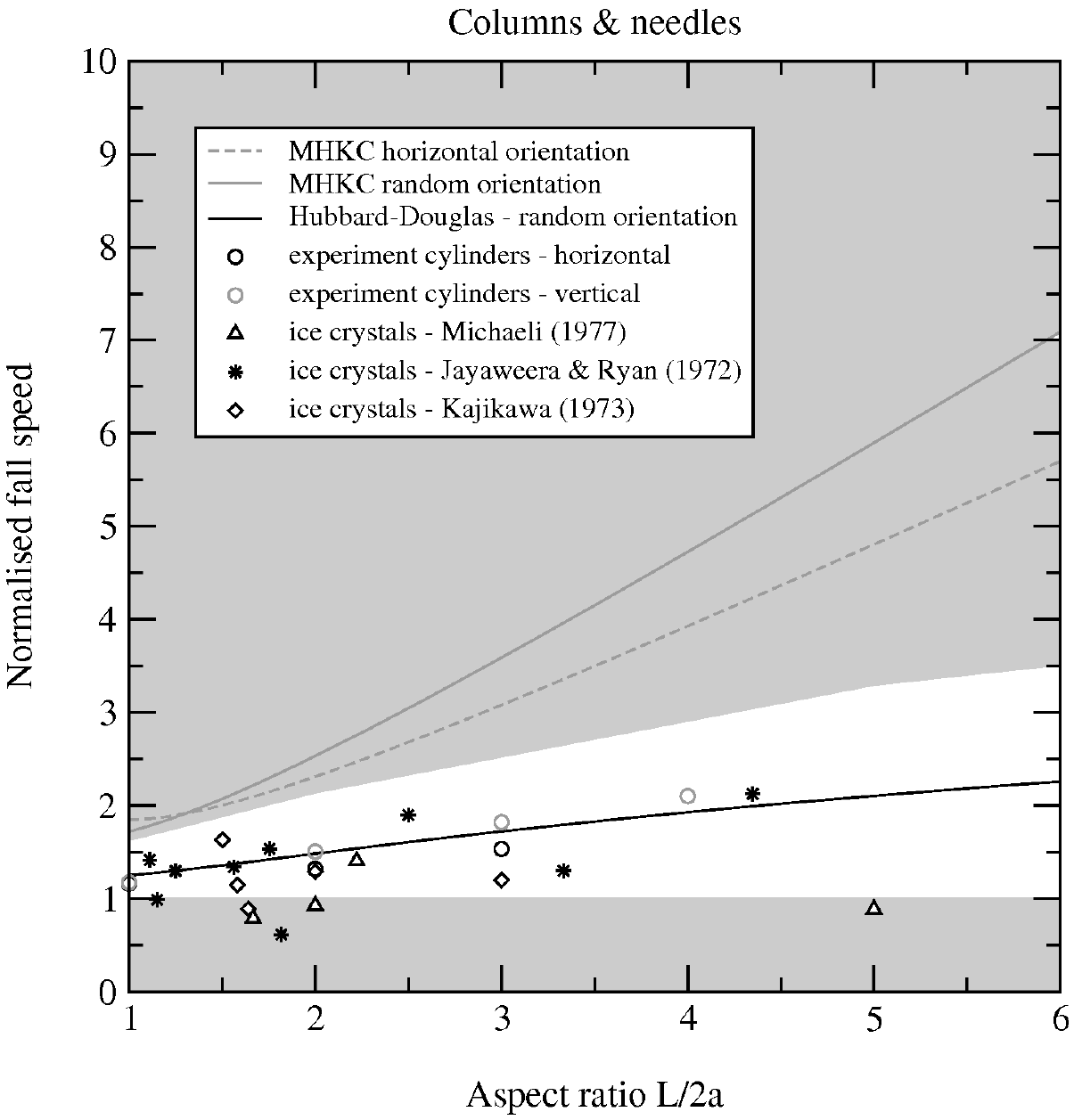}
\caption{\label{columns} Normalised fall speed of hexagonal column and needle ice crystals. Solid black line shows Hubbard-Douglas prediction for randomly oriented crystals ($R=C$). Stars, diamonds and triangles show fall speeds of ice crystals as measured by Jayaweera and Ryan (1972), Kajikawa (1973) and Michaeli (1977) respectively. Circles are measurements of circular cylinders falling horizontally (black circles) and end-on (grey circles). MHKC predictions are shown by the grey solid line (randomly oriented) and grey dashed line (horizontally oriented). Shading indicates fall velocities which lie outside the upper and lower bounds derived from analytical results (see text).}
\end{figure}

Hexagonal column and needle crystals typically grow at temperatures colder than $-22^\circ\mathrm{C}$ and also in a window between $-3$ and $-8^\circ$C (Pruppacher and Klett 1997). Jayaweera and Ryan (1972) made direct measurements of crystal mass, length, width and fall speed; similar measurements were made by Michaeli (1977), and Kajikawa (1973). These data are plotted in figure \ref{columns}, and include solid and hollow hexagonal columns, as well as needles.

The capacitance of a hexagonal column of length $L$ and width $(2a)$ is given by equation \ref{cplate} with $L/2a>1$. Substituting this into the Hubbard and Douglas formula (\ref{hubbard}) yields the black line plotted in figure \ref{columns}. For columns we take the maximum dimension to be $D=\sqrt{L^2+(2a)^2}$. Although there is significant scatter in the experimental data, the bulk of the observations are in agreement with the Hubbard-Douglas curve to within $\sim$20\%. The agreement with experimental data for circular cylinders falling horizontally and end-on (see Happel and Brenner 1965) is also good, to within $\sim$10\%. The cylinders falling end-on have a slightly higher fall speeds than the Hubbard \& Douglas curve (\ref{hubbard}), whilst cylinders falling horizontally have a slightly lower fall speed than (\ref{hubbard}), which seems sensible since the Hubbard-Douglas theory is intended to represent the average over all orientations. 

Also shown in figure \ref{columns} is the MHKC prediction for randomly and horizontally oriented columns. As for planar crystals, both MHKC curves overestimate the experimental data, by around 50\% for an aspect ratio of 1.5, and by more than a factor of two for aspect ratios $L/2a>3$. Upper and lower bounds may be constructed using equal-volume and enclosing spheres respectively as shown by the shaded areas. The MHKC calculations lie above the upper bound for all aspect ratios. 

\subsection{Polycrystals}
Bullet-rosettes are often the dominant crystal habit in cirrus clouds (Heymsfield and Iaquinta 2000) which are typically composed of between two and six bullet crystals in a radial formation. Small `embryonic' bullet-rosettes less than 100$\mu$m in diameter have been observed (see figure 2 of Heymsfield and Iaquinta), and the fall speeds of these particles have been estimated using the Hubbard-Douglas and MHKC theories. There are, to the author's knowledge, no direct observations of the sedimentation velocity of these tiny rosettes.

\begin{figure}
\centering
\includegraphics[width=3in]{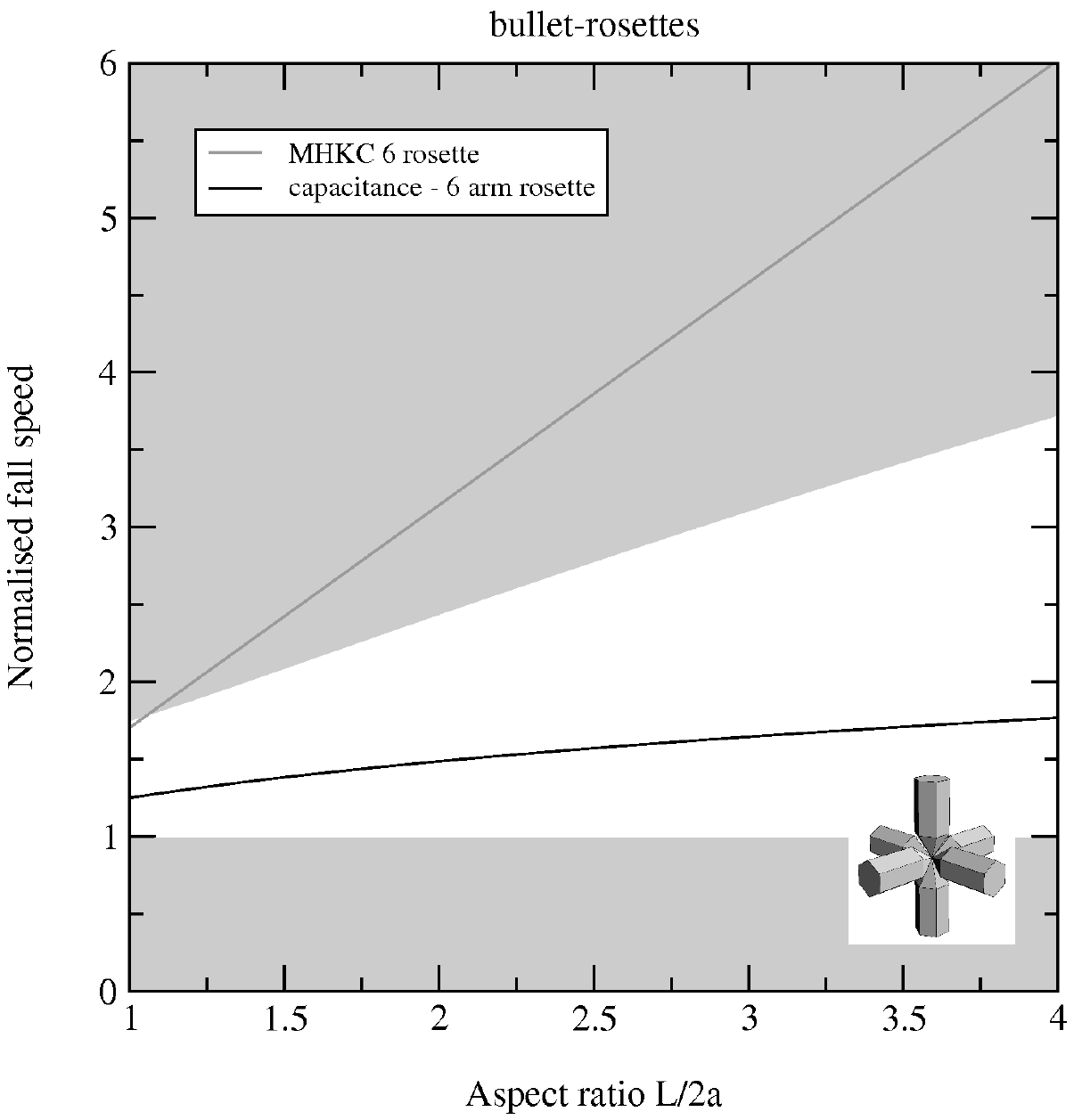}
\caption{\label{rosette} Normalised fall speed of randomly-oriented 6 arm bullet-rosette ice crystals (example shown inset). Black line shows the Hubbard-Douglas prediction as a function of bullet aspect ratio; grey line shows the corresponding MHKC prediction. Shading indicates fall velocities which lie outside the upper and lower bounds derived from analytical results (see text).}
\end{figure}

The capacitance of various bullet-rosette models has been calculated by Westbrook \etal~(2008), who derived the fit $C=0.40D\times(L/2a)^{-0.25}$ for rosettes with six arms (see inset figure \ref{rosette}). Here $L$ is the length of the columnar section of the bullets, and $(2a)$ is their width. The pyramid ends were assumed to be $(L/2)$ in length. Using this data, the fall speeds of bullet-rosettes with a variety of aspect ratios has been calculated using equation \ref{hubbard}, as shown in figure \ref{rosette}.

Also shown in figure \ref{rosette} are the fall speeds of the same model crystals using the MHKC method. As for the capacitance calculations, random orientation is assumed. At aspect ratios close to unity the MHKC fall speeds are $\sim$40\% higher than the capacitance predictions, whilst for rosettes with thin arms $L/2a=4$ the fall speeds are a factor of 3 larger than the Hubbard-Douglas prediction. This implies that the MHKC formula may be significantly overestimating the sedimentation rate of small rosette crystals, especially those with thin arms. This is confirmed by constructing an upper bound using an equal-volume sphere, as indicated by the shaded region.

\section{Discussion}
The sedimentation rate of small ice crystals with maximum dimensions smaller than $\sim100\mu$m have been estimated by exploiting the Hubbard-Douglas (1993) approximation for randomly-oriented particles of arbitrary shape, and the Roscoe (1949) approximation for planar particles settling horizontally. This appears to be the first time (to the author's knowledge) that these approximations have been applied to ice crystals. For hexagonal plate crystals the experimental data lies somewhere between the two approximations suggesting that the crystals were roughly horizontal in their orientation but with some flutter. For dendritic types the experimental scatter was much larger but appears to tell a broadly similar story. For columns and needles the sensitivity to crystal orientation seems to be weaker (given the fall velocities of circular cylinders falling horizontally \textit{vs.} flat on) and the Hubbard-Douglas approximation gives a good estimate for the fall velocity of the crystals. The method has also been applied to model bullet-rosette crystals, for which there are no experimental data to compare with.

The results are in substantial conflict with the MHKC parameterisation, which overestimates the fall speeds of these tiny crystals. From a physical point of view it is perhaps unsurprising that the MHKC boundary-layer formulation fails for $\mathrm{Re}\ll1$. However the suggestion in the past has been that equation \ref{MHKC} would hold down to these small sizes, largely based on the fact that the comparison with Stokes equation for a spherical particle is so good (see section 2). Unfortunately the comparisons with experimental data in this paper show that it does not hold in general for non-spherical ice crystals. Interestingly, Mitchell (1996) did make a comparison with the experiment of Yagi (1970) who observed ice crystals falling in a supercooled fog in Asahikawa, Japan. The average crystal fall speed was $\langle v\rangle=0.107~\mathrm{ms^{-1}}$ and the mean crystal size was close to $100\mu\mathrm{m}$. Mitchell used the recorded habit/diameter distribution, and mass-, area-diameter relationships derived from other studies to estimate $\langle v\rangle=0.097\mathrm{ms^{-1}}$, in apparent agreement. However because the mass/area/size relationships had to be assumed from other studies, the comparison is rather uncertain; also the tiny fall speeds may be strongly affected by even the weakest up/downdraughts present within the fog. The comparisons made in this paper with the direct measurements of individual crystals by Jayaweera and Ryan (1972), Kajikawa (1973) and Michaeli (1977) in controlled laboratory conditions are a much stronger test of the MHKC formula, and one which appears much less favourable to it at small Reynolds number. The upper bound calculations based on analytical solutions for insribed spheroids and equal-volume spheres confirm that equation \ref{MHKC} tends to provide an overestimate for the fall velocity of small crystals settling in a viscous air flow.

It seems that capacitance is a good length scale to correlate ice particle fall speed data for very small crystals. However capacitance is not readily measured from in-situ probes (unlike $A$ and $D$). For the crystal shapes considered here the capacitance may be calculated directly from $D$ using the results in Westbrook \etal~(2008). However there is evidence that many crystals in cirrus clouds may be complex polycrystals of irregular shape (Bailey and Hallett 2002, 2004). Further work is needed to estimate what capacitance such crystals are likely to have. The results given here however highlight the relative insensitivity to crystal shape, with almost all crystal types having a normalised velocity between 1 and 2. This indicates that even for rather tenuous, irregular crystal shapes (such as dendrites) the hydrodynamic radius $R$ is not far removed from circular discs and spheres of the same overall dimensions.

In this paper the focus has been on crystals whose dimensions have been measured explicitly in the lab, and we have found that the MHKC predictions generally overestimated the measured fall velocity. Interestingly, $A-D$ data from aircraft observations can nonetheless produce realistic fall velocities when applied to the MHKC parameterisation in certain circumstances. For example Heymsfield and Miloshevich (2003) measured area ratios of $\gamma\simeq0.75\pm0.15$ using replicator data collected in cold cirrus around $-60^{\circ}$C. This corresponds to normalised velocities in the range 1.3 to 2.0, and much of the experimental data is clustered toward the lower end of this range. The suggestion is that the crystals sampled were quite compact (as evidenced by the relatively high area ratio), a situation in which the MHKC formulation seems to perform best (eg. columns with $L/2a\simeq1$, figure \ref{columns}). For more elongated or tenuous crystals typical of highly supersaturated or mixed-phase conditions, the evidence from figure \ref{plates}, \ref{dendrites}, \ref{columns} and \ref{rosette} is that the MHKC predictions are likely to be rather less realistic.

For $\mathrm{Re}\gg1$ however, the MHKC relationship has proved very successful at accurately predicting ice particle fall speeds, offering good agreement with experimental data for most particle types. The experimental data presented here corresponds to $\mathrm{Re}<0.4$; it is not clear exactly where the cross-over from viscous to boundary-layer flow regimes occurs: in this work I have tentatively suggested a figure of $\sim100\um$ based on the experimental data. More work is needed to determine what happens at the intermediate Reynolds numbers which lie at the boundary between the two regimes. 

The fall speeds of the tiny crystals discussed in this paper are of interest for the interpretation of Doppler lidar measurements in cirrus. It may be possible to assess to what extent small crystals control the optical properties of cirrus cloud by examining the frequency distribution of measured Doppler lidar velocities in such clouds. Using the results from section 2 we find that the maximum fall speed for a $D=60\mu$m crystal is $0.07\mathrm{ms^{-1}}$: this value corresponds to a compact hexagonal column with an aspect ratio of one, and a density equal to that of solid ice. The temperature was assumed to be $-60^{\circ}$C; at warmer temperatures the viscosity of the air is slightly higher and the crystals fall a little slower as a result. Note that the fall speed measurements by Jayaweera and Ryan (1972), Kajikawa (1973) and Michaeli (1977) included crystals over 100$\mu$m in size, and none of them fell faster than $0.06\mathrm{ms^{-1}}$. If the lidar backscatter is dominated by these tiny slow-sedimenting crystals, then the measured mean Doppler velocity should also be dominated by them since it is weighted by the backscatter. Analysis of 1 year of continuous measurements from a 1.5$\mu$m vertically-pointing Doppler lidar at the Chilbolton Facility for Atmospheric and Radio Research in Hampshire is in progress, and the results should help to inform the small ice debate. 

Finally the author notes that the Hubbard-Douglas approximation is likely to be applicable to non-spherical atmospheric aerosol particles, allowing estimates of the sedimentation rate and mobility of complex, irregular aerosols to be made. The only caveat is that the Hubbard and Douglas derivation assumes `stick' boundary conditions, and so may not be suitable for sub-micron particles without an appropriate slip correction.

\section{Appendix A: randomly oriented crystals}
Here the average drag on a randomly oriented ice crystal is estimated using the approach of Hubbard and Douglas (1993). Consider the Oseen tensor, which describes a point hydrodynamic source (Happel and Brenner 1965):
\begin{equation}
\mathbf{T(R)}=\frac{1}{8\pi\eta R}\left(\mathbf{I}+\frac{\mathbf{RR}}{R^2}\right)
\end{equation}
where $\mathbf{I}$ is the identity matrix, and $\mathbf{R=r'-r}$. The orientational average of $\mathbf{T}$ is given by:
\begin{equation}
T=\frac{1}{6\pi\eta R}
\end{equation}
where $R=|\mathbf{R}|$, ie. $T$ is identical to the Green's function for Laplace's equation $(4\pi R)^{-1}$ to within a constant factor. Based on this observation, consider the average flux of momentum away from the ice crystal:
\begin{equation}
\int\sigma T\mathrm{d}s=\frac{1}{6\pi\eta}\times\phi(\mathbf{r})
\end{equation}
where $\mathrm{d}s$ is a small element of surface area, and we define $\sigma$ to be the momentum flux density. Given the above considerations the scalar function $\phi(\mathbf{r})$ must satisfy Laplace's equation, and $\sigma$ may analogously be interpreted as the density of charge on the surface of a conductor with the same size and shape as the ice crystal.

The stress tensor $\mathbf{S}$ is constructed so as to conserve linear and angular momentum, and to ensure that $\mathbf{u_0}$ and the associated drag are co-linear:
\begin{equation}
\mathbf{S}=6\pi\eta\left[(\nabla\phi)\mathbf{u_0}+\mathbf{u_0}(\nabla\phi)-(\nabla\phi)\cdot(\mathbf{u_0})\mathbf{I}\right]
\end{equation}
where $\phi=1$ on the surface of the particle, and $\phi=0$ in the far-field. Then the Stokes equations are:
\begin{equation}
\nabla\cdot\mathbf{S}=6\pi\eta\mathbf{u_0}\nabla^2\phi=0.
\end{equation}
Essentially then, the flow system is described purely in terms of the electrical potential $\phi$ around a conductor of the same size and shape as the ice crystal. The angle-averaged fluid velocity at a given point is $\mathbf{u}(\mathbf{r})=\mathbf{u_0}\phi(\mathbf{r})$.

Given the stress components above, the drag force on the particle is:
\begin{equation}
\left|6\pi\eta\int\mathbf{S}\cdot\mathbf{n}\mathrm{d}s\right|
\end{equation} 
where $\mathbf{n}$ is the unit vector pointing normal to the crystal surface (outwards into the fluid). Using the analogy developed above, the gradient $\nabla\phi$ may be interpreted in terms of the electrical potential gradient near a conductor of the same shape and size as the particle, ie $(\nabla\phi)_{\mathrm{surface}}=-\sigma\mathbf{n}$. The total drag force is then simply:
\begin{equation}
\mathrm{Drag}=6\pi\eta|\mathbf{u_0}|C
\end{equation}
ie. Stokes formula (\ref{stokes}) with $R=C$. 

\section{Appendix B: horizontally oriented planar crystals}
Here the drag on a thin ($L\ll 2a$) planar crystal settling horizontally with velocity $u_0$ is derived, following Roscoe (1949). The Stokes equations are:
\begin{equation}
\eta\nabla^2u=\frac{\partial p}{\partial x}\textrm{, }\eta\nabla^2v=\frac{\partial p}{\partial y}\textrm{, }\eta\nabla^2w=\frac{\partial p}{\partial z}
\label{stokesroscoe}
\end{equation}
where the velocity components $u,v,w$ (corresponding to the $x,y,z$ axes) satisfy the continuity equation:
\begin{equation}
\frac{\partial u}{\partial x}+\frac{\partial v}{\partial y}+\frac{\partial w}{\partial z}=0.
\label{continuity}
\end{equation}
and for convenience we assume $u=v=w=0$ on the crystal surface, and $u=u_0,v=0,w=0$ far from the crystal. The $x$ direction is taken as the vertical axis.

A flow system which satisfies (\ref{continuity}) is:
\begin{equation}
u=\phi-x\frac{\partial\phi}{\partial x}\textrm{, }v=-x\frac{\partial\phi}{\partial y}\textrm{, }w=-x\frac{\partial\phi}{\partial z}
\label{vlap}
\end{equation}
if $\nabla^2\phi=0$ (ie. if $\phi$ is a solution of Laplace's equation). Substituting this into equation \ref{stokesroscoe} the velocities (\ref{vlap}) satisy the Stokes equations provided that:
\begin{equation}
p=-2\eta\frac{\partial\phi}{\partial x}.
\end{equation}
Consider then an earthed conductor with the same shape as the crystal. Taking $\phi=0$ on the surface of the crystal and $\phi=u_0$ far from the crystal matches the boundary conditions on $u,v,w$, and we identify (\ref{vlap}) as the flow system round the ice crystal with $\phi$ the electrical potential around the analogous conductor. 

The normal stress on the surface of the crystal is equal to $p$. The net force normal to the plate is therefore the difference between $p$ on either side of the crystal, ie. $\delta p=2\eta\times 4\pi\sigma$ per unit area of surface (applying Gauss's law), where $\sigma$ is the charge density on the conductor. Integrating $\sigma$ over the whole surface leads to the total translational drag force on the crystal, ie:
\begin{equation}
\mathrm{Drag}=8\pi\eta u_0C
\end{equation}
since the total charge on the conductor is simply the capacitance $C$ multiplied by the applied voltage $u_0$. Comparing this result with Stokes law (\ref{stokes}) leads to the conclusion:
\begin{equation}
R=\frac{4}{3}C
\end{equation}
for planar ice crystals which are sufficiently thin relative to the dimensions of their horizontal cross-section.

\acks{Helpful correspondence with Jack Douglas (NIST) and discussions with Anthony Illingworth and Robin Hogan (Reading) are gratefully acknowledged. The input of two reviewers greatly improved this manuscript. This work was supported by the Natural and Environmental Research Council.

\end{document}